\documentclass{article}
\usepackage{graphicx,geometry}

\begin{document}

\title{Intermediate-spin state and properties of LaCoO$_3$}
\author{M.A.Korotin, S.Yu.Ezhov, I.V.Solovyev, V.I.Anisimov \\
   \it Institute of Metal Physics, Russian Academy of Sciences,\\
   \it 620219 Yekaterinburg GSP-170, Russia \\ \\
       D.I.Khomskii\footnote{Also at the P.N.Lebedev Physical Institute, Moscow, Russia},
       G.A.Sawatzky \\
   \it Solid State Physics Laboratory, University of Groningen,\\
   \it Nijenborg 4, 9747 AG Groningen, \\ The Netherlands}
\date{\today}
\maketitle

\begin{abstract}

The electronic structure of the perovskite LaCoO$_3$ for different spin
states of Co ions was calculated in the LDA+U approach. The ground
state was found to be a nonmagnetic insulator with Co ions in a
low-spin state. Somewhat higher in energy we found two
intermediate-spin states followed by a high-spin state at significantly
higher energy. The calculation results show that Co 3$d$ states of
$t_{2g}$ symmetry form narrow bands which could easily localize whilst
$e_g$ orbitals, due to their strong hybridization with the oxygen 2$p$
states, form a broad $\sigma^*$ band. With the increase of temperature
which is simulated by the corresponding increase of the lattice
parameter, the transition from the low- to intermediate-spin states
occurs. This intermediate-spin (occupation $t_{2g}^5e_g^1$) can develop
an orbital ordering which can account for the nonmetallic nature of
LaCoO$_3$ at 90~K$<$T$<$500~K. Possible explanations of the magnetic
behavior and gradual insulating-metal transition are suggested.
\end{abstract}

\section{INTRODUCTION}

Among the systems showing semiconductor to metal transition LaCoO$_3$
is especially interesting due to the fact that is also displays very
unusual magnetic behavior, often associated with low-spin (LS) --
high-spin (HS) transition \cite{Good}. Although a large number of
investigations have been carried out since the early 1960's the
character of the transition and the nature of the temperature
dependence of the spin state is still unclear. For example the
temperature dependence of the magnetic susceptibility shows a strong
maximum at around 90~K followed by a Curie-Weiss-like decrease at
higher temperatures \cite{Asai} which was interpreted by the authors as
a LS to a HS transition. The semiconductor to metal-like transition
occurs in a range of 400 -- 600~K, well above this transition. The
insulating nature of LaCoO$_3$ below 400 -- 600~K was attributed by
Raccah and Goodenough to an ordering of LS and HS Co$^{3+}$ ions in a
NaCl like structure \cite{Good}, with the itinerant electrons in a
broad band formed by the transition metal $e_g$ orbitals. This
structural change however has not been observed in crystallographic
studies. Recent neutron scattering experiments \cite{Neutron} suggest
that the semiconductor -- metal transition is not predominantly
magnetic in origin but a clear picture of what its nature is absent.

Photoemission (XPS) and X-ray absorption (XAS) studies on the other
hand have been interpreted in terms of a spin change at the
semiconductor -- metal transition \cite{Abbspec}. There are however
several discrepancies in the interpretation. The Co 2$p$ XAS line shape
at room temperature looks like LS Co$^{3+}$ which is not consistent
with a spin state transition at 90~K. However, the valence band XPS
spectrum at 300~K is quite different from that of a HS compound. So the
interpretation of the two transitions is still uncertain.

In most of the previous studies one has assumed a rather ionic and
ligand field-like starting point. In this picture the possibility of
intermediate spin (IS) states as well as local symmetry lowering,
orbital ordering and the large bandwidth of the $e_g$ states were not
considered. It has recently been pointed out by several authors that
the oxides corresponding to high formal oxidation states may be
negative charge transfer systems in the Zaanen-Sawatzky-Allen scheme
\cite{ZSA} resulting in an essential modification of the electronic
structure, in particular in a possible stabilization of an IS states
\cite{PotzeSCO}.

A first physical model to explain the transitions in LaCoO$_3$ was
proposed and recently revised by Goodenough in \cite{Good}. It was
suggested that for trivalent cobalt in LaCoO$_3$ the crystal field
energy $\Delta_{cf}$ is only slightly larger than the intraatomic
(Hund) exchange energy $\Delta_{ex}$.  Thus the ground state of the Co
ion is LS $^1A_1$ (S=0) and the excited HS state $^5T_2$ (S=2),
according to \cite{Good}, is only 0.08~eV higher in energy. The
corresponding one-electron configurations are $t_{2g}^6e_g^0$ for the
LS state and $t_{2g}^4e_g^2$ for the HS state. The increase in
temperature leads to the population of the HS states which is reflected
in the magnetic susceptibility measurements. The semiconductor-metal
transition was interpreted in a rather vague way as a formation of a
$\sigma^*$ band from the localized ionic $e_g$ states. Long-range order
of the LS and HS Co ions was assumed based on the results of the X-ray
diffraction measurements. The recent neutron diffraction experiments
\cite{Neutron} did not confirm it, but did not exclude a short-range
order.

The theoretical description of LaCoO$_3$ is a difficult task, because this
system exhibits transition from the localized to the delocalized
behavior, and the two main existing methods: one-electron band
structure calculations based on the Local Density Approximation (LDA)
and "model hamiltonian configuration interaction" approach (which were
both used for the compound under consideration \cite{Abbspec,Abbband})
were constructed for the, respectively, completely itinerant and fully
localized cases.

Recently a paper of Sarma {\it et al} was published \cite{sarma} where
the results of the LSDA calculations for La$M$O$_3$ ($M$=Mn, Fe, Co, Ni)
were presented. It was claimed there that LSDA results gave good
agreement with the X-ray photoemission spectra. Such an agreement is
not necessary a proof of the adequate description of the electronic
structure. A well known example is NiO, where LSDA gives a sharp peak
at the top of the valence band in agreement with XPS, but the nature of
this peak is Ni3d $t_{2g}$ minority-spin character, while it is
generally accepted that this peak is derived from $d^8\underline{L}$
final state. The real proof of the correctness of the one-electron
approximation would be an absence of the satellites in the
photoemission spectra. Also the XPS spectra, with which a comparison
was made in \cite{sarma}, were measured at the room temperature, where
LaCoO$_3$ is already not in the low-spin state. So XPS must be compared
with LSDA result corresponding to the magnetic solution and not a
non-magnetic one as in \cite{sarma}.

Anisimov {\it et al} proposed a so-called LDA+U method \cite{Anis} which
combines in one calculation scheme LDA and Hubbard model approaches and
is able to treat on a "first-principle" basis the systems with strong
Coulomb correlation. It was demonstrated that this method (in contrast
to standard LDA) could describe the existence of the different states
of the system under consideration which are close in total energy
(holes in doped copper oxides \cite{PRL} and transition metal
impurities in insulators \cite{mgo}).

In this paper we present a study of an electronic structure of
LaCoO$_3$ in the LDA+U approach. In contrast to the standard LDA there
are several stable solutions corresponding to different local minima of
the LDA+U functional. We have found a nonmagnetic insulating ground
state in agreement with the experiment. We have also found two
orbitally-polarized magnetic solutions corresponding to IS states (one
of them is gapless semiconductor and the other is metal) and
semiconducting magnetic solution corresponding to a HS state of the
system, which lies much higher in energy.  In addition we found that
the $e_g$ states form broad bands whereas the $t_{2g}$ states exhibit
narrow bands, split by Coulomb interactions into the lower (occupied)
and upper (unoccupied) Hubbard bands. Using the results of these
calculations, we propose an interpretation of the behavior of
LaCoO$_3$. According to our scheme, there first occurs with increasing
temperature a transition from a LS (nonmagnetic) insulating ground
state to a state with an IS (configuration $t_{2g}^5e_g^1$). Due to
strong Jahn-Teller nature of this configuration, this state may develop
orbital ordering. The orbitally-ordered state turns out to be
nonmetallic (actually nearly zero-gap semiconductor) in our
calculations. With the further increase of the temperature the orbital
ordering may be gradually destroyed which can explain the transition to
a metallic state observed in LaCoO$_3$ at 400 -- 600~K.  We hope that
this new information will eventually lead to a better understanding of
LaCoO$_3$ and other similar materials.

\section{CALCULATION METHOD}

The LDA+U method \cite{Anis} is based on the assumption that it is
possible to separate all electrons in the system in two subsets:  the
localized states (for LaCoO$_3$ these are the Co 3$d$-orbitals), for
which Coulomb intrashell interactions are described by the Hubbard-like
term, and the itinerant states, where the averaged LDA energy and
potentials are good approximations.

It is known that LDA calculations can provide all the necessary model
parameters (such as the Coulomb parameter $U$ \cite{agun}, the exchange
$J$, hopping parameters describing hybridization {\it etc.}) on a
first-principle basis, but the one-electron structure of the LDA
equations with the orbital-independent potential does not allow to use
these parameters in the full variational space. LDA+U overcomes this
problem by using the framework of the degenerate Anderson model in the
mean-field approximation. In this method, the trial function is chosen
as a single Slater determinant, so it is still a one-electron method,
but the potential becomes orbital-dependent, and that allows one to
reproduce the main features of strongly correlated systems: the
splitting of the $d$-band to the occupied lower Hubbard band and
unoccupied upper Hubbard band.

The main idea of our LDA+U method is that the LDA gives a good
approximation for the average Coulomb energy of $d$-$d$ interactions
$E_{av}$ as a function of the total number of $d$-electrons
$N=\sum_{m\sigma} n_{m\sigma}$, where $n_{m\sigma}$ is the occupancy of
a particular $d_{m\sigma}$-orbital:

\begin{equation}
E_{av}=U\frac{N(N-1)}{2}-J\frac{N(N-2)}{4}.
\end{equation}

But LDA does not properly describe the full Coulomb and exchange
interactions between $d$-electrons in the same $d$-shell. So we
suggested to subtract $E_{av}$ from the LDA total energy functional and
to add orbital- and spin-dependent contributions to obtain the exact
(in the mean-field approximation) formula:

\begin{equation}
E=E_{LDA}-\bigl(U\frac{N(N-1)}{2}-J\frac{N(N-2)}{4}\bigr)
    +\frac{1}{2}\sum_{m,m^\prime,\sigma}U_{mm^\prime}
n_{m\sigma}n_{m^\prime -\sigma}
    +\frac{1}{2}\sum_{m\neq m^\prime,m^\prime,\sigma}
(U_{mm^\prime}-J_{mm^\prime})n_{m\sigma}n_{m^\prime \sigma} \; .
\end{equation}

The derivative of Eq.(2) over orbital occupancy $n_{m\sigma}$ gives the
expression for the orbital-dependent one-electron potential:

\begin{equation}
V_{m\sigma}(\vec{r})=V_{LDA}(\vec{r})+
 \sum_{m^\prime}(U_{mm^\prime}-U_{eff})n_{m^\prime -\sigma}
 +\sum_{m^\prime \neq m}(U_{mm^\prime}-J_{mm^\prime}
 -U_{eff})n_{m\sigma}
 +U_{eff}(\frac{1}{2}-n_{m\sigma})-\frac{1}{4}J \;.
\end{equation}

The Coulomb and exchange matrices $U_{mm^\prime}$ and $J_{mm^\prime}$
are:

\begin{eqnarray}
U_{mm^\prime} & = & \sum_{k}a_{k}F^{k} , \\
J_{mm^\prime} & = & \sum_{k}b_{k}F^{k} , \\
a_{k} & = & \frac{4\pi}{2k+1}\sum_{q=-k}^{k} \langle lm|Y_{kq}|lm\rangle
\langle m^\prime|Y_{kq}^\ast|l m^\prime\rangle , \\
b_{k} & = & \frac{4 \pi}{2k+1}\sum_{q=-k}^{k}
|\langle lm|Y_{kq}|l m^\prime\rangle |^2 \; ,
\end{eqnarray}

where the $F^{k}$ are Slater integrals and $\langle lm|Y_{kq}|l
m^\prime\rangle $ are integrals over products of three spherical
harmonics $Y_{lm}$.

For $d$-electrons, one needs $F^0, F^2$ and $F^4$ and these can be
linked to the parameters $U$ (direct Coulomb interaction) and $J$
(intraatomic exchange) obtained from the LSDA-supercell procedures
\cite{agun} via $U=F^0$ and $J=(F^2+F^4)/14$, while the ratio $F^2/F^4$
is to a good accuracy constant $\sim 0.625$ for 3$d$ elements
\cite{groot}. For LaCoO$_3$, the Coulomb parameter U was found to be
7.8~eV and the exchange parameter J=0.92~eV.

The LDA+U approximation was applied to the full potential linearized
muffin-tin orbitals (FP-LMTO) calculation scheme \cite{FPLMTO}.
Crystallographic data being used in the calculations were taken from
\cite{Neutron}. According to them, LaCoO$_3$ has a pseudocubic
perovskite structure with a rhombohedral distortion along the (111)
direction. The unit cell contains two formula units. Since this
rhombohedral distortion is small (the largest rhombohedral angle is
60.990$^\circ$ at 4~K), we use the conception of $t_{2g}$ and $e_g$
orbitals, as referred to in the cubic setting in the following
discussion. Temperature was introduced in our calculations only by the
change of lattice parameter and atomic positions according to the data
of Ref.\cite{Neutron}. The most detailed description of the technical
aspects of FP~LMTO calculations for the perovskite-type complex oxides
can be found in \cite{Andrei}. The optimal choice of the basis set for
describing the valence band and the bottom of conduction band of the
LaCoO$_3$ is presented in Table~\ref{basisset}. Since U-correction is
applied for the Co $d$-orbitals, the value of the {\it muffin-tin} (MT)
radius for Co was chosen close to its value in metallic Co in order to
get the full 3$d$-density inside the sphere. For the correct
description of the wave functions in the interstitial region, we
expanded the spherical harmonics up to the value of l$_{max}$=5,~4,~3
for La, Co and O MT-spheres correspondingly. The Brillouin zone (BZ)
integration in the course of the self-consistency iterations was
performed over a mesh of 65 {\bf k}-points in the irreducible part of
the BZ. Densities of states (DOS) were calculated by the tetrahedron
method with 729 {\bf k}-points in the whole BZ.
\begin{table}
\centering
\caption{Basis set and MT-sphere radii used in the calculation.}
\label{basisset}
\begin{tabular} {llllc}
\hline
\multicolumn{1}{l}{Atom}
&\multicolumn{1}{c}{$\kappa^2_1=-0.01$ Ry}
&\multicolumn{1}{c}{$\kappa^2_2=-1.0$ Ry}
&\multicolumn{1}{c}{$\kappa^2_3=-2.3$ Ry}
&\multicolumn{1}{c}{$R_{MT},$~\AA}\\
\hline
La &6$s$6$p$5$d$4$f$ &6$s$6$p$5$d$4$f$ &6$s$6$p$5$d$ &1.77\\
Co &4$s$4$p$3$d$     &4$s$4$p$3$d$     &4$s$4$p$     &1.26\\
O  &2$s$2$p$         &2$s$2$p$         &2$s$         &0.66\\
\hline
\end{tabular}
\end{table}

As the potentials for the various $d$-orbitals of Co are different in
LDA+U, it is not obvious {\it a~priori} what will be the final symmetry
of the system under consideration. In preliminary calculations we
assumed for the simplicity that LaCoO$_3$ has a perfect perovskite-type
cubic lattice. To allow the system to choose the appropriate symmetry
by itself, we perform an integration not over 1/48 part of BZ as for
cubic symmetry but over 1/8 (D$_{2h}$ symmetry group). The resulting
symmetries were found to be cubic O$_{h}$ for the ground (LS) state and
tetragonal D$_{4h}$ for the excited states. Practically the same
situation occurs in real R$\bar{3}$c symmetry: the occupancies of $xy,
yz, zx$ orbitals are about the same for the low-spin configuration as
well as that of $3z^2-r^2, x^2-y^2$ orbitals. The degeneracy of $xy$
and $yz, zx$ orbitals is broken in the other spin states.

\section {RESULTS AND DISCUSSION}
\subsection {Homogeneous Solutions}

We start by considering the results of the calculations for homogeneous
regimes, without extra superstructure. The possibility to get a
solution with an orbital ordering is considered in section \ref{sec:orbord}.

The detailed results of the calculations with crystallographic data
corresponding to 4~K are presented in Table \ref{tab:res}. One must
start with the equal occupancies of all three $t_{2g}$ orbitals and
also of two $e_g$ orbitals (e.g. $t_{2g}^6e_g^0$) to obtain the LS
configuration. The initial spin polarization vanished during
self-consistency iterations resulting in a nonmagnetic final solution.
The charge excitation spectrum has a semiconducting character in
accordance with experimental data \cite{Gap} (and in contrast to the
LDA result \cite{Abbband}) with the energy gap equal to 2.06~eV. The
top of the valence band (Fig.1) is formed by the mixture of oxygen 2$p$
states with Co $t_{2g}$ orbitals and the bottom of conduction band
predominantly by the $e_g$ orbitals.

\begin{table}
\centering
\caption{Total energy difference, character of energy spectrum,
occupancies of various $d$-orbitals of Co and magnetic moments
for various Co spin-states in LaCoO$_3$.}
\label{tab:res}
\begin{tabular}{cllclllcccc}
\hline
\multicolumn{1}{c}{configu-}
&\multicolumn{1}{c}{$\delta$ $^a$}
&\multicolumn{1}{c}{energy}
&\multicolumn{7}{c}{d-occupancies}
&\multicolumn{1}{c}{$\mu_{d-Co}$}\\
\multicolumn{1}{c}{ration}
&\multicolumn{1}{c}{(eV)}
&\multicolumn{1}{c}{gap (eV)}&
&\multicolumn{1}{c}{$xy$}
&\multicolumn{1}{c}{$yz$}
&\multicolumn{1}{c}{$zx$}
&\multicolumn{1}{c}{$3z^2-r^2$}
&\multicolumn{1}{c}{$x^2-y^2$}
&\multicolumn{1}{c}{total}
&\multicolumn{1}{c}{($\mu_B$)}\\
\hline
$t_{2g}^6e_g^0$& --& 2.06& $\uparrow$,$\downarrow$
                 & 0.98& 0.98& 0.99& 0.32& 0.33& 7.20& 0\\
low-spin\\
\hline
$t_{2g}^5\sigma^*$ & 0.24& 0 $^b$ &  $\uparrow$
                 & 0.98& 0.98& 0.98& 0.84& 0.84& 7.13& 2.11\\
intermediate-spin& & & $\downarrow$ & 0.07& 0.98& 0.99& 0.23& 0.24\\
\hline
$t_{2g}^4e_g^2$& 0.65& 0.13&  $\uparrow$
                 & 0.99& 0.99& 0.99& 1.00& 1.00& 6.78& 3.16\\
high-spin&&& $\downarrow$& 0.99& 0.08& 0.11& 0.33& 0.30\\
\hline
\multicolumn{11}{l} {$^a$ Total energy difference relative to the energy of
$t_{2g}^6e_g^0$ configuration.} \\
\multicolumn{11}{l} {$^b$ Half-metal. Total magnetic moment per unit cell
is 2 $\mu_B$.}\\
\hline
\end{tabular}
\end{table}

\begin{figure}
  \centering
  \includegraphics[width=0.6\textwidth]{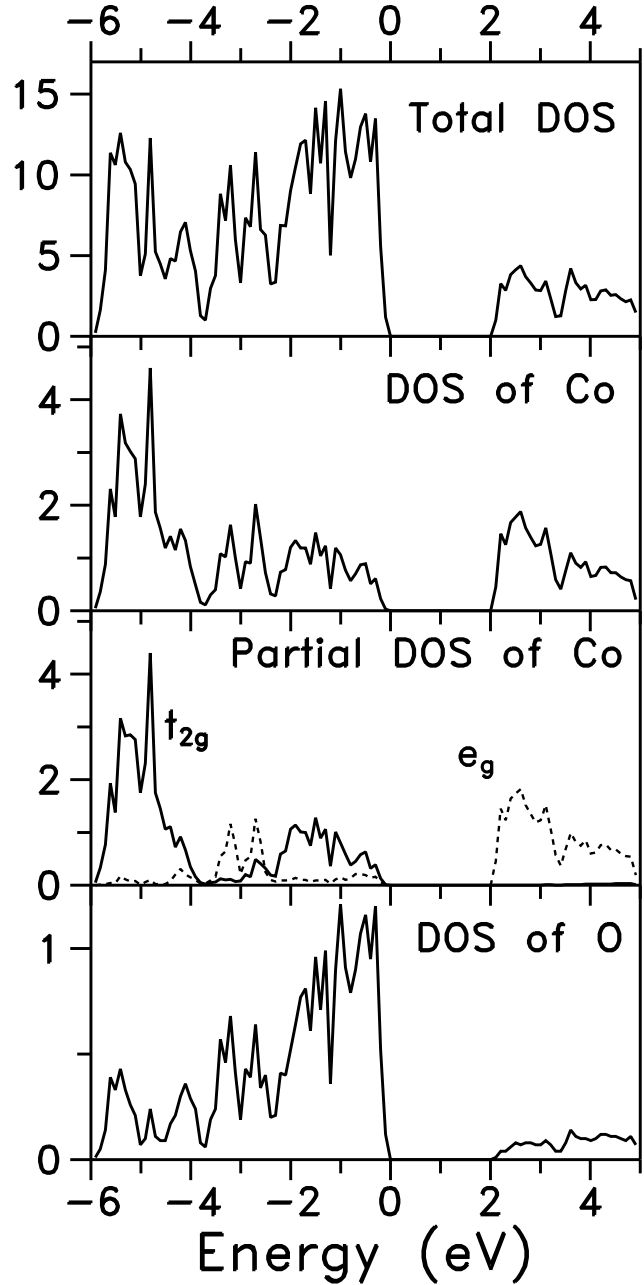}
  \caption{The total and partial densities of states obtained in LDA+U
   calculations for LaCoO$_3$ with Co ions in low-spin state
   ($t_{2g}^6e_g^0$ configuration). For $d$-Co partial density of states
   solid line denotes $t_{2g}$ orbitals and dashed line -- $e_g$ orbital.}
   \label{Fig.1}
\end{figure}

From the X-ray photoemission and absorption spectra \cite{Abbspec} the
gap was estimated to be 0.9~$\pm$~0.3~eV. The optical measurements
\cite{arima} gave $\approx$~0.5~eV. The larger value of the calculated
gap can be explained by the well known fact that a mean-field
approximation, which is the basis of our LDA+U approach, usually
overestimates the tendency to localization and hence the values of the
gap.

The aim of our work was to find solutions corresponding to the
higher-spin states of Co ions. Usually the HS state is described as a
$t_{2g}^4e_g^2$ configuration with the maximum value of spin S=2
(magnetic moment $\mu_{Co}$=4~$\mu_B$). This corresponds to the purely
ionic model, and hybridization of the Co 3$d$-orbitals with the oxygen
2$p$-orbitals and band formation in a solid can renormalize
significantly this ionic value. Such kind of renormalization was
obtained in our calculations (see Table \ref {tab:res}). The initial
$t_{2g}^4e_g^2$ configuration (two holes on $d_{xz}^\downarrow$,
$d_{yz}^\downarrow$ orbitals of the $t_{2g}$ set and two on
$d_{3z^2-r^2}^\downarrow$, $d_{x^2-y^2}^\downarrow$ of the $e_g$ set
with spin-down (minority) spin projection) results in the
self-consistent solution with a magnetic moment of
$\mu_{d-Co}$=3.16~$\mu_B$. The total energy of this HS solution is
0.65~eV per formula unit higher than the ground state LS
$t_{2g}^6e_g^0$ configuration. The unexpected result is that there
exists an {\it intermediate-spin} solution (second line in Table
\ref{tab:res}, magnetic moment value $\mu_{d-Co}$=2.11~$\mu_B$) which
is lower in total energy than the HS solution (0.24~eV per formula unit
relative to the LS ground state). This solution was obtained when we
started from the initial configuration $t_{2g}^5e_g^1$, where only one
electron was transferred from the $t_{2g}$ to $e_g$ states. The final
self-consistent configuration is better described as a $t_{2g}^5$ state
with a partially filed $\sigma^*$-band ($t_{2g}^5\sigma^*$) with the
occupancies of the $d_{3z^2-r^2}^\uparrow$ and $d_{x^2-y^2}^\uparrow$
orbitals equal to 0.84. In a configuration interaction language used in
the cluster calculations this may be compared to a
$d^6$+$d^7\underline{L}$ state where $\underline{L}$ denotes a hole on
the oxygen.

The IS state in our calculations turns out to be metallic (see,
however, section \ref{sec:orbord}), and HS state is semiconducting
(Fig.2,3). The reason for this is that the antibonding $\sigma^*$ band
(formed by $e_g$ orbitals) is very broad and the band splitting is not
strong enough to create a gap in the case of IS state. When the
$e_g^\uparrow$ band becomes completely filled (HS state), a small gap
between $e_g^\uparrow$ and e$_g^\downarrow$ bands appears.
\begin{figure}
  \centering
  \includegraphics[width=0.7\textwidth]{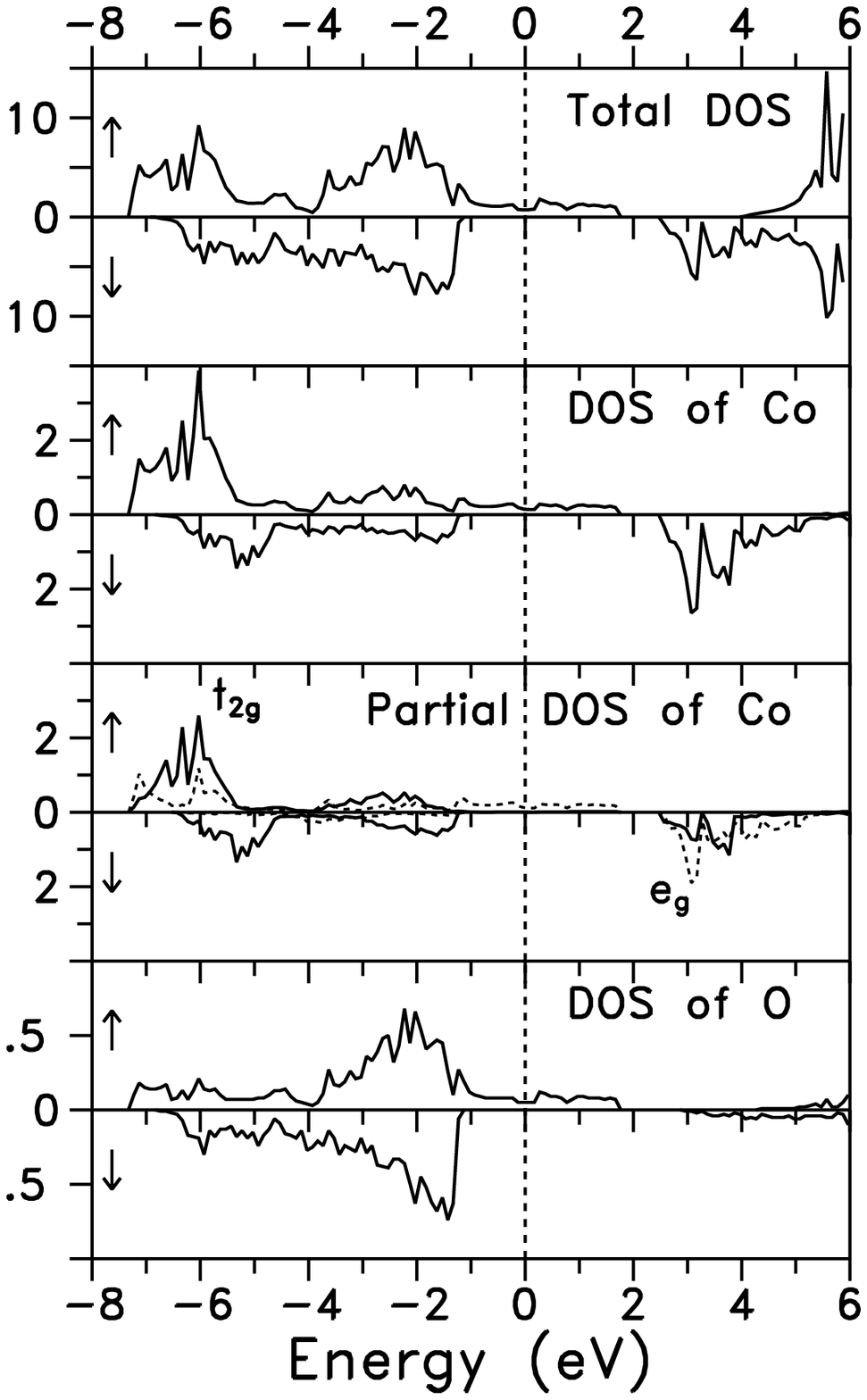}
  \caption{The total and partial densities of states obtained in LDA+U
   calculations for LaCoO$_3$ with Co ions in intermediate-spin state
   ($t_{2g}^5\sigma^*$ configuration). Fermi level is denoted by vertical
   dashed line. Arrows corresponds to spin-up and spin-down spin projection.
   The other denotations are the same as on Fig.1.}
  \label{Fig.2}
\end{figure}

In Table \ref{tab:res} the occupation of different orbitals in
different configurations is presented. One notices that the actual
occupation is different from the formal "chemical" one and corresponds
on the average not to 6 but rather to 7 electrons in $d$-shell. The
reason for this is the strong hybridization of the empty $e_g$ orbitals
with oxygen 2$p$-orbitals. In the LS ground state every $e_g$ orbital,
which is formally empty, has actually an occupancy of about 0.33
resulting in 1.3 additional $d$-electrons above the formal six-electron
configuration. In the case of excited IS state we have partially filled
broad $\sigma^*$-band with the total number of $e_g$ electrons
increased by 0.85 as compared to the LS state and the number of
$t_{2g}$-electrons 0.92 less than in LS state with as a result
practically unchanged total number of $d$-electrons. Therefore, we used
the notation $t_{2g}^5\sigma^*$ in the Table \ref{tab:res} to prevent
misunderstanding and in accordance with notation proposed in
\cite{Good}. One may also say that despite the formal oxidation state
Co$^{3+}$ the real configuration e.g. in the IS state is a mixture of
$t_{2g}^5e_g^1$ and $t_{2g}^5e_g^2\underline{L}$ configurations.

\begin{figure}
 \centering
 \includegraphics[width=0.7\textwidth]{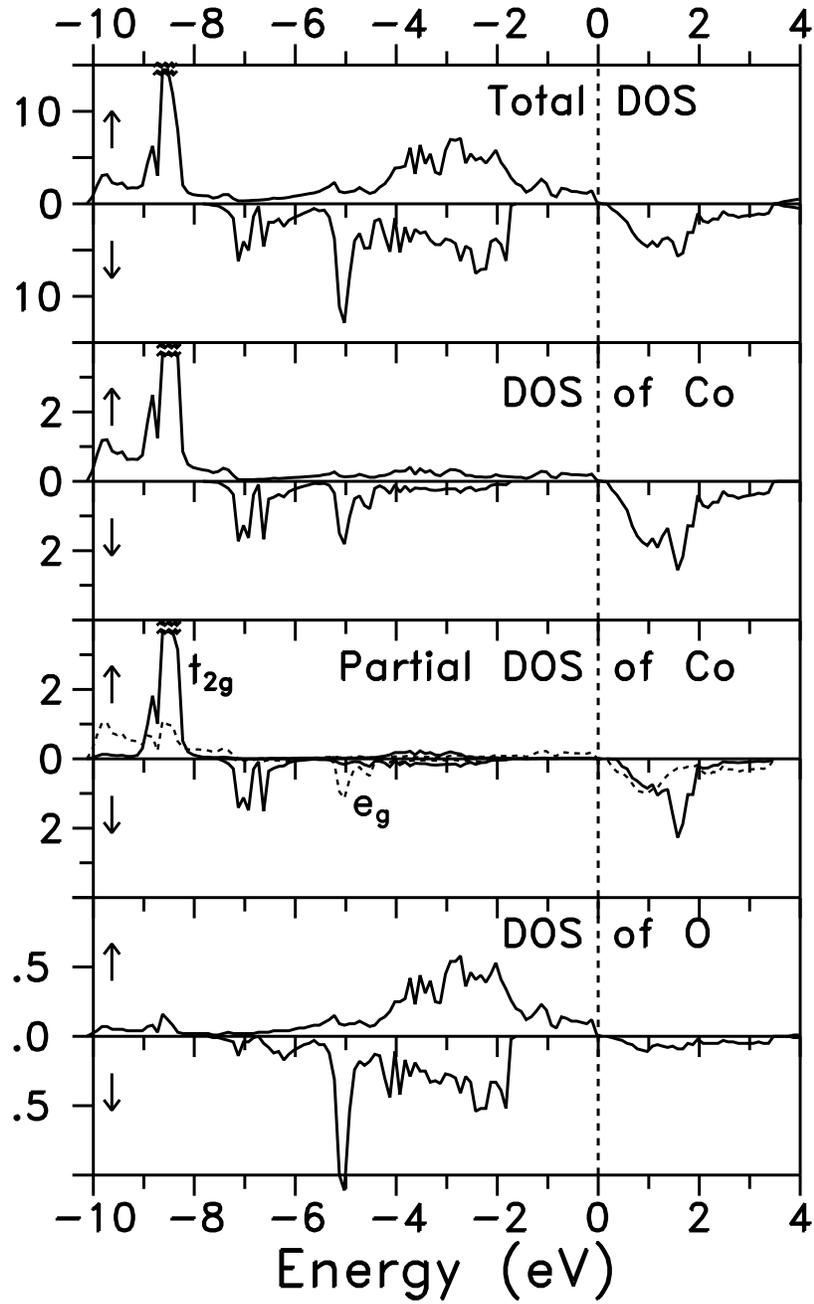}
 \caption{The total and partial densities of states obtained in LDA+U
 calculations for LaCoO$_3$ with Co ions in high-spin state
 ($t_{2g}^4e_g^2$ configuration).  Denotations are the same as on
 Fig.1 and Fig.2.}
 \label{Fig.3}
\end{figure}

Some other points concerning the results obtained are worth mentioning.
As one can see from Fig.1, the photoemission holes in a LS state are
mainly formed by oxygen states (note that in Figs.1-3 the partial
densities of states of constituent atoms are given per one atom, and
there are 3 of oxygen atoms per formula unit). Consequently the final
state of the XPS for LS state formally corresponds to
d$^6\underline{L}$ configuration. At the same time the electronic
excitations mostly reside on Co $e_g$ orbitals hybridized with
$p$-states of oxygen. However, in the IS case which in our calculations
is metallic (Fig.2), the states at the Fermi level contain comparable
weight of both Co $e_g$ and O $p$ orbitals.

The IS state obtained is fully polarized which is due to the
half-metallic ferromagnetic nature of the solution: the magnetic moment
per formula unit is 2~$\mu_B$, i.e. it corresponds to the spin S=1.  In
that sense the calculation agrees with the ionic picture, in which the
IS configuration of Co is d$^6$ ($t_{2g}^5e_g^1$), with S=1.

The IS state of LaCoO$_3$ in our results turned out to be half-metal
\cite{hmgroot} ferromagnet, in which only the electrons of one spin
projection are at the Fermi level (the other spin subbands are fully
occupied). Ferromagnetic arrangement of the magnetic moments may be of
course an artifact of the scheme used. In principle, LDA+U method is
intended to describe the local correlation effects such as formation of
the Hubbard gaps and local magnetic moments, which should still be
present in the paramagnetic phase. However technically in the
calculations we have to assume certain long-range magnetic order in
accordance with translation symmetry of the crystal.  What the results
obtained in such a way do really tell you, is whether there is a gap or
not and what is the value of the {\it local} magnetic moment.

Although the IS state in the calculations turns out to be metallic, the
density of states at the Fermi level is very low, n(E$_F$)=0.36 states
per eV per one formula unit. This probably indicates that it would not
be so difficult to make this state semiconducting, see the discussion
below (section \ref{sec:orbord}).

The results presented above show that the first excited configuration
-- that with the IS -- lies only 0.24~eV higher than the ground state
with the LS. Experimentally it is established that there is a
transition from LS nonmagnetic state to the magnetic one with the
increase of temperature. According to the Ref.\cite{Asai} this
transition occurs in the vicinity of 90~K. As the closest magnetic
state is that with the IS (HS state lies, according to our results,
much higher at 0.65~eV), we ascribe nonmagnetic -- magnetic transition
in LaCoO$_3$ to the LS -- IS transition.

It is well known (see, e.g., \cite{Good}) that Co in the HS state has
larger ionic radii than in LS one, and LS -- HS transitions are
accompanied by the increase of the volume (and vice versa). Keeping
that in mind we carried out the calculations of the electronic
structure of LaCoO$_3$ at different lattice parameters which can
imitate the influence of the temperature (via thermal expansion).
Fig.4 demonstrates the values of total energies for various spin states
of Co relative to the energy of $t_{2g}^6e_g^0$ state at the lattice
parameter corresponding to 4~K. In this figure actual lattice
parameters used in calculations are shown on the horizontal axis
together with the temperature scale (we used the lattice parameters as
a function of the temperature measured in Ref.\cite{Neutron}). One sees
that with the increase of the specific volume (or with the increase of
temperature) the energy of the IS state crosses that of the LS state,
which corresponds to the LS -- IS transition. In our calculation this
crossover occurs at about 150~K -- somewhat larger that the
experimental value about 90~K. Nevertheless this transition temperature
is correct by order of magnitude which for such {\it ab~initio}
calculations is rather satisfactory. Note that the HS state still lies
high enough even at the temperatures about 600~K although we can not
exclude that it could have become the ground state at still larger
specific volumes or temperatures.

\begin{figure}
 \centering
 \includegraphics[width=\textwidth]{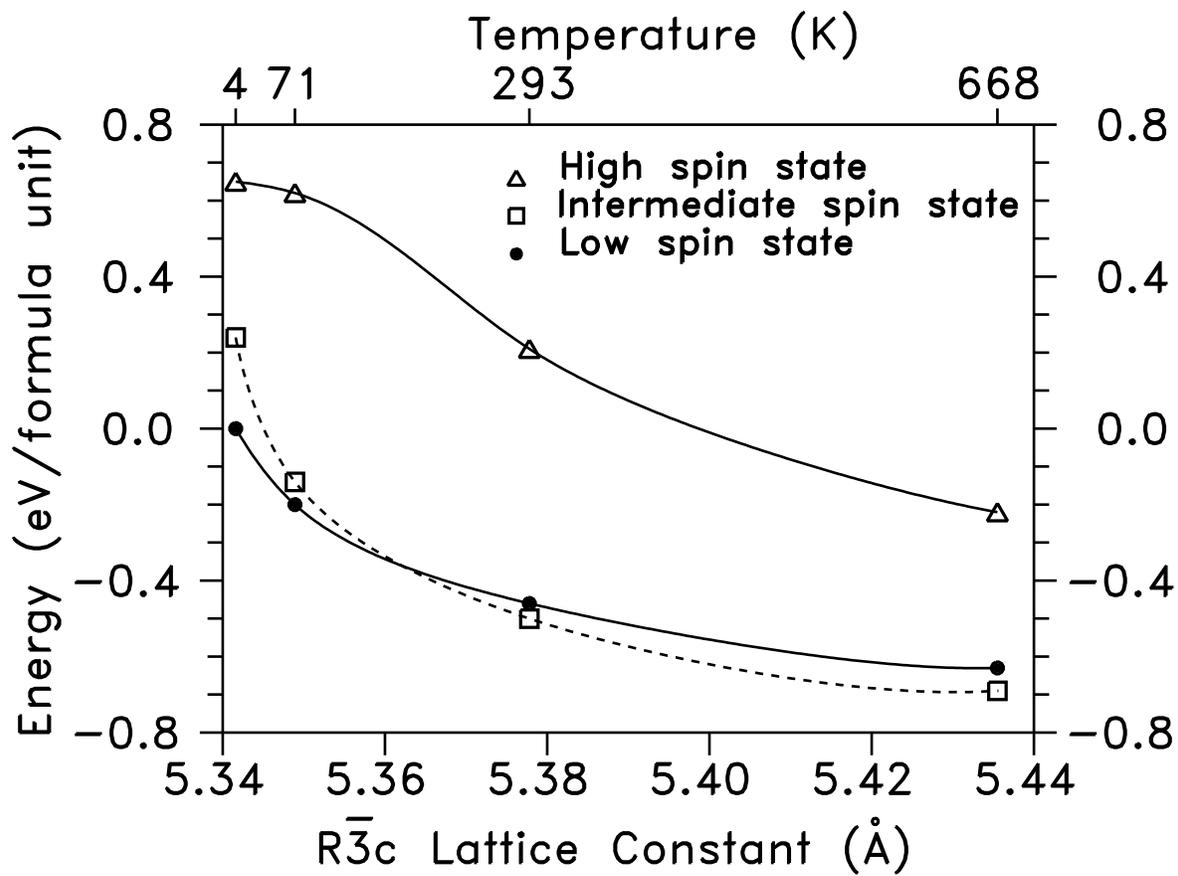}
 \caption{The total energies for various spin states of LaCoO$_3$ relative
 to the energy of $t_{2g}^6e_g^0$ state at 4~K versus R$\bar{3}$c lattice
 constant. The correspondent temperatures are marked also.}
 \label{Fig.4}
\end{figure}

The total energy for all three solutions (LS, IS, and HS) decreases
with the increasing volume and minima are achieved only for lattice
parameters corresponding to high temperature. At this value of the volume
the total energy of the IS state is lightly lower than LS. It is well known
that the equilibrium values of volume calculated with LSDA are always
a few percent less than the experimental volume, because LSDA overestimates
cohesion. LSDA+U, on the other hand, underestimates cohesion due to its treating
the $d$ states  as localized. In both cases the calculated lattice parameters
significantly deviate from the experimental values. Goodenough suggested that
the magnetic transition in LaCoO$_3$ is caused by the fact that the crystal-field energy $\delta{_cf}$ is only slightly larger than the intra-atomic (Hund)
exchange energy $\delta{_ex}$.  The crystal-field energy is determined by the
3$d$-2$p$ hopping parameters, which strongly depend on crystal volume. Hence if
one will perform calculations with lattice parameters corresponding to the
equilibrium volume, then the delicate balance between the $\delta{_cf}$ and
$\delta{_ex}$. Our results show that for low-temperature lattice parameters
is larger $\delta{_cf}$ than the $\delta{_ex}$, but already a small increase
on volume (corresponding to increasing the temperature to 150 K) would reverse
this ratio.

\begin{figure}
 \centering
 \includegraphics[width=\textwidth]{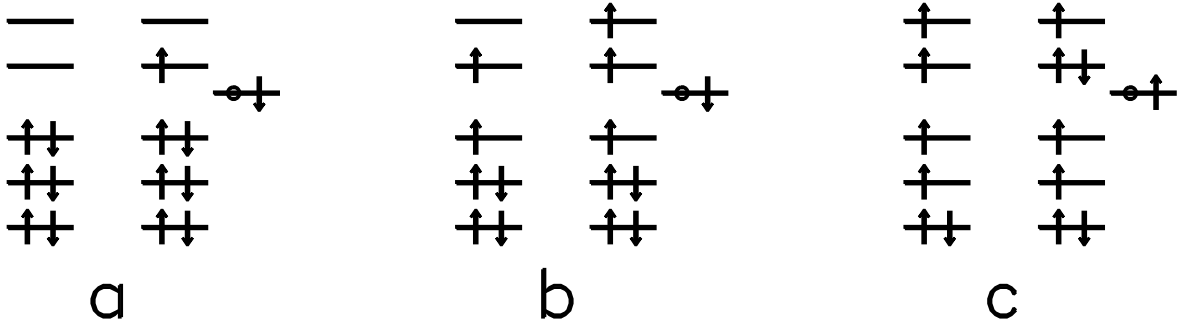}
 \caption{Scheme representation of various Co $d^6$+$d^7\underline{L}$
  configurations in low- ({\bf a}), intermediate- ({\bf b}) and high-
  ({\bf c}) spin states. Open circle denotes a hole in oxygen $p$-shell.}
 \label{Fig.5}
\end{figure}

Thus from the results obtained we conclude that the nonmagnetic --
magnetic transition in LaCoO$_3$ most probably occurs not between LS
and HS states, but between {\it low}- and {\it intermediate}-spin
states. The reason for the stabilization of the IS state may be
understood from the following consideration. Different configurations
of Co$^{3+}$ ($d^6$) are illustrated in Fig.5.  Here $t_{2g}$ and $e_g$
are atomic $d$-states splitted by the crystal field (the splitting
$\Delta$=10~Dq). In purely ionic picture one should expect that,
depending on the ratio between $\Delta$ and intraatomic exchange
interaction J either LS or HS states should be stable (if the energy of
the LS state E$_{LS}$ is taken to be zero, then E$_{IS}$=$\Delta$-J and
E$_{HS}$=2$\Delta$-6J, so that LS state is the ground state for
$\Delta>$3J and HS state -- if $\Delta<$3J; IS state in this scheme
always would lie higher). However, as we mentioned in the Introduction,
the oxides with unusually high valence of transition metals have a
tendency to have the $d$-shell occupancy corresponding to a lower
valence, with the extra hole being predominantly located on oxygen (one
gets a configuration $d^7\underline{L}$ instead of $d^6$). In this
situation the $d$-$p$ hybridization is especially strong and plays a
crucial role. These considerations are also supported by our
calculations, see Table \ref{tab:res}.

Especially important is the hybridization with the $e_g$ orbitals. One
can show that it is stronger for the IS state than for the HS one:
there are more channels of $p$-$e_g$ hybridization in this case (the
most favorable configurations of the type $d^7\underline{L}$, admixed
to $d^6$, are also shown in Fig.5). In other words, if these
configurations ($d^7\underline{L}$) would be dominant, then the state
corresponding to the ground state of the $d^7$ configuration would have
lowest energy -- and it is known that for $d^7$ (Co$^{2+}$) it is just
the configuration of Fig.5b -- i.e., the one giving a total state of
IS. That is the physical reason why the IS configuration lies below the
HS one and may become the ground state for an expanded lattice (higher
temperatures).

The nonmagnetic -- magnetic transition occurring at $\sim$90~K which we
now associate with the LS -- IS transition, is described quite
reasonably by our model. The transition temperature obtained in our
calculations ($\sim$150~K) lies not so far from that experimentally
observed. In Ref.\cite{Asai} the relative change of the lattice
parameter ($\Delta a$) at this transition was fitted by the relation
(see \cite{Ches})

\begin{equation}
\Delta a = \frac{\nu*exp(-E_q/kT)}{1+\nu*exp(-E_q/kT)},
\end{equation}

where $E_q$ is the energy gap (it was taken in \cite{Asai} as a fitting
parameter), and $\nu$ is degeneracy of the magnetic state. As the
high-temperature magnetic ground state was believed to be the HS state
in Ref.\cite{Asai}, the degeneracy was taken $\nu$=15 (triply
degenerated $t_{2g}$ orbital times the spin degeneracy (2S+1) with S=2,
see Fig.5c). In our new interpretation the degeneracy will be $\nu$=18
(triply degenerated $t_{2g}$ orbital times double degeneracy of $e_g$
orbital times (2S+1) with S=1, see Fig.5b). Thus the fit of $\Delta a$
would be as successful with this interpretation as with the one given
in \cite{Asai}, with actually nearly the same value of the energy gap
$E_q$.

The picture of an IS state can also help to resolve one more problem
mentioned in Ref.\cite{Asai}: that the correlations between magnetic
sites in LaCoO$_3$ are not antiferromagnetic but (weakly)
ferromagnetic.  One should expect only antiferromagnetic correlations
between {\it high-spin} Co$^{3+}$ ions in which the $e_g$-shell is
half-filled ($e_g^2$). However in IS Co ions have nominally $e_g^1$
configuration, and especially if these $e_g$ orbitals are ordered (see
discussion below) one can have a ferromagnetic exchange interaction
(the well-known example is e.g. ferromagnetic K$_2$CuF$_4$ \cite{KKh}).

\subsection {Possible Orbital Ordering in LaCoO$_3$}
\label{sec:orbord}

The conception of LS -- IS transition describes quite reasonably the
nonmagnetic -- magnetic transition observed in LaCoO$_3$. At the same
time it was established that the first transition at $\sim$90~K leaves
this compound semiconducting whereas our calculated IS has a metallic
character of the energy spectrum. Is it possible to overcome this
disagreement?

As is clear from Fig.5b, the IS state has a strong double-degeneracy
($e_g^1$ configuration). For localized electrons it would be the
typical Jahn-Teller situation. In particular for KCuF$_3$ it was
proposed in Ref.\cite{KKh} and it was confirmed by calculations in
Ref.\cite{Licht} that there exist the special kind of orbital and
magnetic ordering of $d$-ions in simple cubic lattice with one electron
or hole on doubly-degenerate $e_g$ level. The situation in IS LaCoO$_3$
is very similar to the case of KCuF$_3$ (keeping in mind that there
exist permanent rhombohedral distortion and there are already two Co
ions in an elementary cell in LaCoO$_3$, the "antiferro" orbital
ordering may be consistent with the existing structural data).

We checked this possibility by repeating the calculations, assuming now
that there exist an orbital ordering. The structure of the
corresponding ordered IS state was taken as consisting of ferromagnetic
planes (001) whereas the direction of spins is opposite in the nearest
planes. The occupied $e_g$ orbitals were assumed to alternate; as a
starting point the orbitals in two sublattices were taken as
$d_{x^2-z^2}$ and $d_{y^2-z^2}$, see Fig.6. To supplement this Figure
with the information about the other $d$-orbitals, let's consider the
Co ion in $t_{2g}^5e_g^1$ configuration which has
$d_{y^2-z^2}^\uparrow$ occupied $e_g$ orbital (corresponding site is
marked as {\bf 1} on Fig.6). For this state the other three $e_g$
orbitals are empty and all the $t_{2g}$ orbitals are occupied, with the
exclusion of $d_{yz}^\downarrow$ one. This choice minimizes the Coulomb
interaction energy of $t_{2g}$ and $e_g$ electrons. For the
neighboring Co ion (site {\bf 2} in Fig.6) the $e_g$ electron is
placed in $d_{x^2-z^2}^\uparrow$ orbital and $t_{2g}$ hole is placed in
$d_{xz}^\downarrow$ orbital. Calculations were made for the real
rhombohedral crystal structure with lattice parameters corresponding to
71~K. With this state as a starting point the self-consistent
calculation was now carried out. The resulting densities of states
corresponding to such orbital ordered IS state are presented in Fig.7.

\begin{figure}
 \centering
 \includegraphics[width=\textwidth]{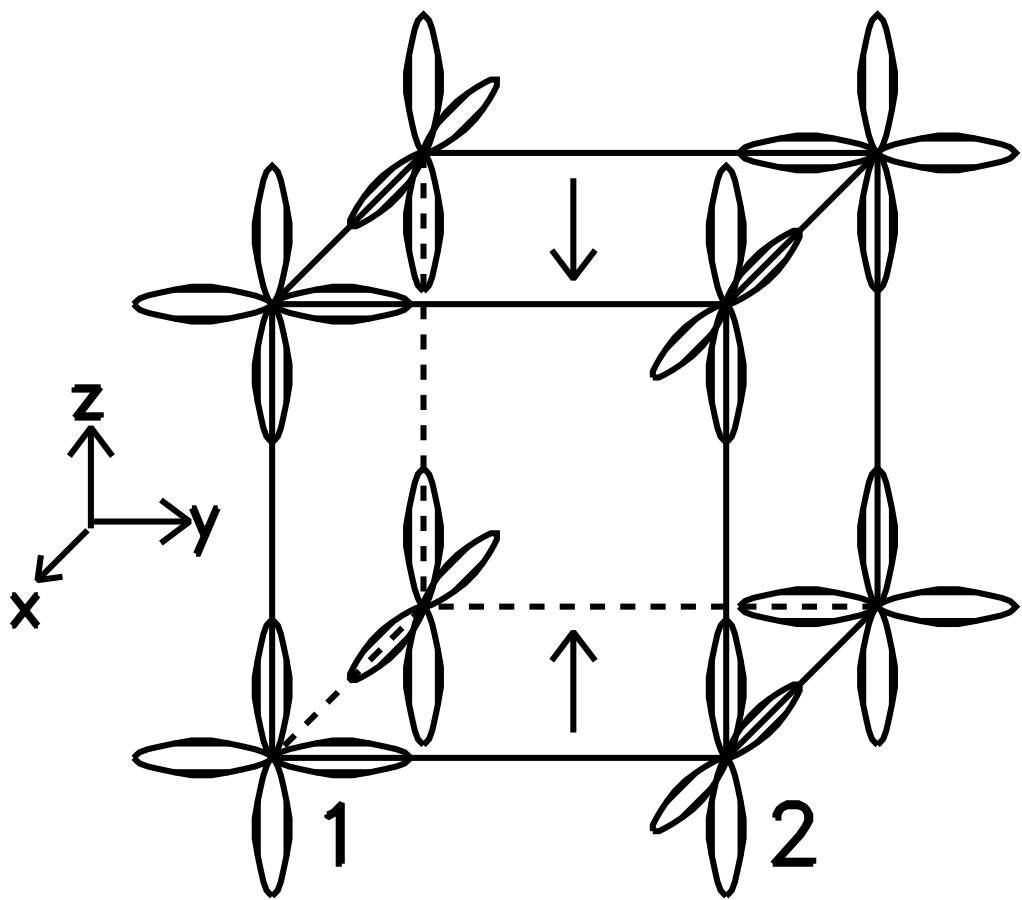}
 \caption{Spin and orbital ordering in orbital ordered intermediate spin
  state for occupied $e_g$ orbitals. For the simplicity the perfect cubic
  structure for Co ions is shown.}
 \label{Fig.6}
\end{figure}

First of all it is necessary to point out that such kind of spin and
orbital ordering is stable with respect to the self-consistency
iterations and hence there exists the solution with such symmetry. The
character of an energy spectrum turned  out to be gapless
semiconducting-like. The wide $e_g$ band which crossed the Fermi level
in the case of IS state without orbital order (see Fig.2) is now
splitted. The top of the valence band is formed (for one sublattice,
e.g. for Co site {\bf 1}) by $d_{y^2-z^2}$ states and the bottom of
conduction band -- by $d_{3x^2-r^2}$ states (Fig.7b,c) hybridized with
the oxygen states. As to the $t_{2g}$ densities of states, the
character of their energy distribution is practically not changed in
comparison with the case of IS state without orbital ordering. The
value of spin magnetic moment of Co is 1.87~$\mu_B$ in this orbitally
ordered IS state and it is less than for one without orbital order
(2.11~$\mu_B$).

\begin{figure}
 \centering
 \includegraphics[width=0.6\textwidth]{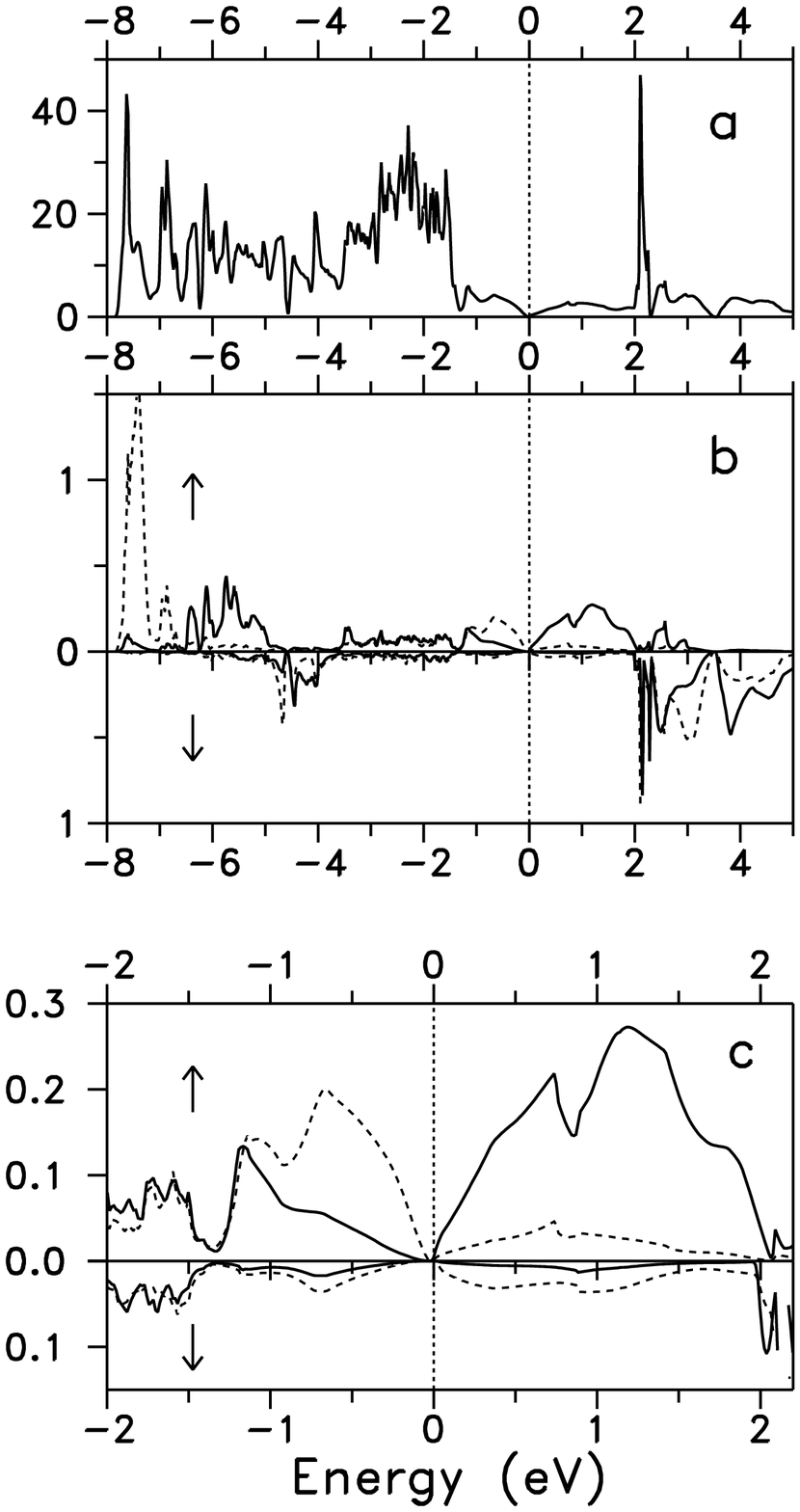}
 \caption{The total and partial densities of states obtained in LDA+U
 calculations for LaCoO$_3$ with Co ions in ordered intermediate-spin
 state.
 {\bf a}: Total density of states (per 4 formula units and both spins);
 {\bf b}: partial Co $e_g$ densities of states e.g. for Co site {\bf 1}
 in Fig.6: solid line - for $d_{y^2-z^2}$ states, dashed line - for
 $d_{3x^2-r^2}$ states;
 {\bf c}: the same as in {\bf b} in the vicinity of the Fermi level.
 Denotations are the same as on Fig.2.}
 \label{Fig.7}
\end{figure}

When we compare the two IS states with the fixed spin magnetic moment
1.87~$\mu_B$, the total energy of the orbitally ordered IS state is
0.11~eV lower. The necessity of using a fixed magnetic moment is due
to the limitation of the band structure scheme. The magnetic state of
LaCoO$_3$ is known to be paramagnetic (with the possible short-range
spin or orbital ordering). At the same time as we already mentioned
above, it is necessary to set a long-range magnetic order in
calculations (antiferromagnetic for orbitally ordered IS state and
ferromagnetic for one without orbital order). This long-range magnetic
ordering imposed (in addition to the intraatomic exchange) influences
somehow both the local magnetic moment and the total energy. In the
case of long-range antiferromagnetic order the value of the magnetic
moment is apparently underestimated in comparison with the real
paramagnetic state, for the ferromagnetic one -- overestimated. Thus
to estimate the relative stability of different phases we compare the
energies calculated for the same value of the magnetic moment. Our
calculations show that in the case of equal spin values the orbitally
ordered IS state is more preferable than the one without orbital order.

One can interpret now the first nonmagnetic -- magnetic transition (at
about 90~K) as a transition of Co ions from LS to IS state with the
specific orbital ordering of occupied $e_g$ orbitals. Our results show
that this transition occurs at temperatures lower than 150~K. Under
this transition a magnetic moment on Co sites appears whereas the
material is still nonmetallic. The spin-only value of the {\it
effective} magnetic moment, in the case of IS state (S=1), is
$\mu_{eff}$=2.83~$\mu_B$ -- somewhat below experimental value of
3.1--3.4~$\mu_B$. It will be increased by the orbital contribution,
because there is in principle an unquenched moment of $t_{2g}$
electrons for the configuration $t_{2g}^5e_g^1$ (which e.g. for
Co$^{2+}$ ions the increases the value of $\mu_{eff}$ typically by
$\sim$0.3~$\mu_B$). We cannot treat this effect numerically because our
codes do not account for the spin-orbit interaction; however we expect
on physical grounds that this effect should be present (which, by the
way, would change somewhat the occupation of $t_{2g}$ orbitals).  An
extra change of $\mu_{eff}$ may be due to the very strong $d$-$p$
hybridization which leads to a large contribution to the total wave
function of a state $d^7\underline{L}$ ($t_{2g}^5e_g^2\underline{L}$).
In this state the effective moment of Co ion is that of the high-spin
Co$^{2+}$ (S=3/2) compensated by the opposite polarization of the
oxygen $p$-shell (S=-1/2) (note that this picture is supported by the
calculations (see Table \ref{tab:res}) where local Co spin moment is
2.11~$\mu_B$ and the total spin moment per unit cell is 2~$\mu_B$).
This fact may significantly change the value of $\mu_{eff}$ measured
in the susceptibility experiments.

The IS state with the nominal configuration $t_{2g}^5e_g^1$ may have
an orbital ordering because of the doubly degenerate $e_g$ orbitals
(strong Jahn-Teller nature of this configuration). This turned out to
be the case in our calculations: the orbitally ordered state may be
crudely described as an alternation of the occupied $d_{z^2-x^2}$ and
$d_{z^2-y^2}$ orbitals.

According to our calculations, this ordering leads to the splitting of
the $e_g$ ($\sigma^*$)-band, leading practically to the zero-gap
situation, with the Fermi level lying in the gap. The second gradual
transition to the metallic state ($\sim$600~K) with the increase of the
effective magnetic moment value up to $\mu_{eff}$=4.0~$\mu_B$ may then
be associated with the disappearance of the orbital ordering with
increased temperature still within the IS state. This IS state without
orbital order is calculated to be a metal with a larger magnetic
moment. Our calculations show that the second transition has no
relations to the HS state of Co ions because the HS solution lies still
high in energy and if realized would have been semiconductor.

\section{CONCLUSION}

The calculations of the electronic structure in LDA+U approximation for
LaCoO$_3$ were performed. At 4~K the lowest total energy has the
nonmagnetic insulating solution with Co ions in LS state. Three excited
state configurations (two IS and one HS) are also found to be stable,
with local magnetic moments on Co sites. The $t_{2g}$ states exhibit
narrow bands whilst the $e_g$ states form broad bands. With the
increase of lattice parameter corresponding to the thermal lattice
expansion two transitions can be expected: the first occurs from LS
state to IS state with the orbital ordering which in our calculation is
a zero-gap semiconductor. The second transition occurs within the IS
state and is connected with gradual disorder of occupied $e_g$
orbitals.  We believe that although many details are not clear yet, the
concept of the nonmagnetic -- magnetic and semiconductor -- metal
transitions in LaCoO$_3$ as connected mainly with IS states is a good
candidate to explain the properties of this interesting material which
remains a puzzle for so long.

We thank Dr.A.Postnikov and R.Potze for helpful discussions. This work
was supported by the Netherlands Foundation for Fundamental Research on
Matter (FOM), the Netherlands Foundation for Chemical Research (SON),
the Netherlands Organization for the advancement of Pure Research
(NWO), the Committee for the European Development of Science and
Technology (CODEST) program and the Netherlands NWO special fund for
scientists from the former Soviet Union.

\end{document}